\documentclass[11pt]{article}%
\usepackage{geometry}
\usepackage{dsfont}
\usepackage{amsmath}
\usepackage{amsfonts}
\usepackage{amssymb}
\usepackage{graphicx}%
\usepackage{subfig}

\begin{document}

\date{}
\title{\textbf{Supersymmetric generalization of the\\ maximal Abelian gauge}}
\author{\textbf{M.~A.~L.~Capri}$^{a}$\thanks{caprimarcio@gmail.com}\,\,,
\textbf{H.~C.~Toledo}$^{a}$\thanks{henriqcouto@gmail.com}\,\,,
\textbf{J.~A.~Helay\"el-Neto}$^{b}$\thanks{helayel@cbpf.br}\,\,\\[2mm]
{\small \textnormal{$^{a}$  \it Departamento de F\'{\i }sica Te\'{o}rica, Instituto de F\'{\i }sica, Universidade do Estado do Rio de Janeiro, UERJ -}}
 \\ \small \textnormal{\phantom{$^{a}$} \it Rua S\~{a}o Francisco Xavier 524, 20550-013, Rio de Janeiro, RJ, Brasil}\\
	\small \textnormal{$^{b}$ \it Centro Brasileiro de Pesquisas F\'{\i}sicas, CBPF - Rua Dr Xavier Sigaud 150, 22290-180, Rio de Janeiro, RJ, Brasil}\normalsize}
\maketitle
\begin{abstract}
We present an $N=1$ supersymmetric generalization of the Maximal Abelian Gauge (MAG) in its superfield and component-field formulations, and discuss its Faddeev-Popov quantization, the associated BRST symmetry and the corresponding Slavnov-Taylor identity in superspace approach.
\end{abstract}

\baselineskip=13pt



\maketitle
\section{Introduction}
In ordinary (non-supersymmetric) Yang-Mills theories, the Maximal Abelian Gauge (MAG) is a non-linear gauge that enjoys some interesting properties,
such as renormalizability \cite{'tHooft:1981ht,Min:1985bx,Fazio:2001rm}. It naturally splits the diagonal and off-diagonal components of the gauge field, $\mathcal{A}_{\mu}(x)$, being then suitable
for the study of the so-called Abelian dominance conjecture, according to which the infrared limit of the Yang-Mills theories might be governed only by
diagonal degrees of freedom \cite{Ezawa:1982bf}. In Euclidean space, the MAG can be defined in the lattice, providing several numerical results about the behavior of the gluon
propagator in the infrared regime \cite{Mendes:2006kc,Bornyakov:2003ee,Mihari:2007zz}, and it also allows the study of the Gribov ambiguity problem \cite{Capri:2005tj,Capri:2010an,Capri:2008ak}, a feature which, according to our knowledge, is also well established only in
the Landau gauge\footnote{An example of how to deal with the Gribov ambiguity problem in the linear covariant gauges is given by \cite{Sobreiro:2005vn}. However, it seems to be possible only for small values of the gauge parameter $\alpha$, {\it i.e.} very close to the Landau gauge, which corresponds to $\alpha=0$.}, thanks to the Hermiticity of Faddeev-Popov operator\footnote{The Gribov ambiguity problem was originally proposed in the Landau gauge in \cite{Gribov:1977wm}; extended by Zwanziger in \cite{Zwanziger:1989mf,Zwanziger:1992qr}, giving rise to the so-called Gribov-Zwanziger (GZ) model; and refined by Dudal {\it et al} in \cite{Dudal:2008sp},  giving rise to the refined Gribov-Zwanziger (RGZ) model. The GZ and RGZ scenarios correspond to a very complex and extensive chapter, which is not already concluded, in the understanding of the behavior of the Yang-Mills theories in the infrared. We strongly recommend for a complete review the references \cite{Sobreiro:2005ec,Vandersickel:2012tz,Vandersickel:2011zc}. }.\\\\In $N=1$, $D=4$ super-Yang-Mills (SYM) theory, the Gribov problem also takes place and it has recently received
some attention in the supersymmetric version of Landau gauge \cite{Amaral:2013uya} (in a superfield formalism) and  in Landau-Wess-Zumino gauge \cite{Capri:2014xea} (in a component formalism).
We recognize that such applications should be reproduced in a supersymmetric version of the MAG. Then, in this work, we would like to present, as a first step,
the supersymmetric version of the MAG in the superfield formalism and discuss the Faddeev-Popov quantization as well as the BRST symmetry and the Slavnov-Taylor
Ward identity. Also we are able to recover the ordinary MAG in the component formalism by imposing the Wess-Zumino gauge conditions.\\\\The paper is then
organized as follows: in Section2, we briefly review the ordinary MAG and present its supersymmetric generalization. In Section~3, we discuss the
Faddeev-Popov quantization and the BRST symmetry. We also present here the Slavnov-Taylor identity, which is fundamental for renormalization purposes \cite{Piguet:1995er}.
Finally, in Section~4, we display our conclusions and perspectives for future works.

\section{The ordinary and the supersymmetric versions of the MAG}
\subsection{The diagonal and off-diagonal generators of the SU($n$) group}
Let us start by expanding the gauge field $\mathcal{A}_{\mu}(x)$ in a base of generators of the SU($n$) group:
\begin{equation}
\mathcal{A}_{\mu}(x)=A^{A}_{\mu}(x)T^{A}=A^{a}_{\mu}(x)T^{a}+A^{i}_{\mu}(x)T^{i}\,.
\end{equation}
The set of the $(n^{2}-1)$ generators $T^{A}$ can be divided into two subsets: one subset is composed by the $n(n-1)$ off-diagonal generators,
represented here by $T^{a}$, and the other one is composed by the $(n-1)$ diagonal generators, represented here by $T^{i}$. Then, we shall adopt
the following notation: the indices $\{A,B,C,D\dots\}$ run from $1$ to $(n^{2}-1)$; the off-diagonal indices $\{a,b,c,d,e,\dots\}$ run from $1$ to
$n(n-1)$; and the diagonal indices $\{i,j,k,l,\dots\}$ run from $1$ to $(n-1)$.\\\\Now, taking into account the diagonal and off-diagonal sectors of
the group and following the notation described above, the Lie algebra of the generators is given by\footnote{These relations can be obtained directly
from $\big[\,T^{A},T^{B}\,\big]=if^{ABC}T^{C}$.}:
\begin{equation}
\big[\,T^{a},T^{b}\,\big]=if^{abc}T^{c}+if^{abi}T^{i}\,,\qquad
\big[\,T^{a},T^{i}\,\big]=-if^{abi}T^{b}\,,\qquad
\big[\,T^{i},T^{j}\,\big]=0\,,
\label{lie}
\end{equation}
where $f^{abc}$ and $f^{abi}$ are the structure constants of the group, being completely antisymmetric under successive permutations of the indices
and obeying the following useful identities\footnote{These identities can be obtained directly from the Jacobi identity:\,$
f^{ABC}f^{CDE}+f^{ADC}f^{CEB}+f^{AEC}f^{CBD}=0$.}:
\begin{eqnarray}
0&=&f^{abi}f^{bcj}+f^{abj}f^{bic}\,,\nonumber\\
0&=&f^{abc}f^{cdi}+f^{adc}f^{cib}+f^{aic}f^{cbd}\,,\nonumber\\
0&=&f^{abc}f^{cde}+f^{abi}f^{ide}+f^{adc}f^{ced}+f^{adi}f^{ieb}
+f^{aec}f^{cbd}+f^{aei}f^{ibd}\,.
\end{eqnarray}
Also, the commutation relation between the diagonal generators, the third equation of \eqref{lie}, shows that there is an Abelian, U$(1)^{n-1}$,
subgroup of SU($n$). Actually, this subgroup is known as the Cartan subgroup \cite{Georgi:1999wka}.\\\\The field-strengths, $\mathcal{F}_{\mu\nu}(x)$,
of the gauge field can also be decomposed in terms of the diagonal and off-diagonal components:
\begin{equation}
\mathcal{F}_{\mu\nu}(x)=F^{a}_{\mu\nu}(x)T^{a}+F^{i}_{\mu\nu}(x)T^{i}\,,
\end{equation}
where\footnote{The relations \eqref{Fs} can also be obtained from $F^{A}_{\mu\nu}=\partial_{\mu}A^{A}_{\nu}-\partial_{\nu}A^{A}_{\mu}
+f^{ABC}A^{B}_{\mu}A^{C}_{\nu}$.}
\begin{eqnarray}
F^{a}_{\mu\nu}&=&D^{ab}_{\mu}A^{b}_{\nu}-D^{ab}_{\nu}A^{b}_{\mu}+f^{abc}A^{a}_{\mu}A^{b}_{\nu}\,,\nonumber\\
F^{i}_{\mu\nu}&=&\partial_{\mu}A^{i}_{\nu}-\partial_{\nu}A^{i}_{\mu}+f^{abi}A^{a}_{\mu}A^{b}_{\nu}\,,\nonumber\\
D^{ab}_{\mu}&=&\delta^{ab}\partial_{\mu}-f^{abi}A^{i}_{\mu}\,,
\label{Fs}
\end{eqnarray}
with $D^{ab}_{\mu}$ being the covariant derivative with respect to the diagonal components.\\\\
Thus, the ordinary Yang-Mills action can be written as:
\begin{equation}
S_{\mathrm{YM}}=-\frac{1}{2g^{2}}\,\mathrm{tr}\int d^{4}x\,\mathcal{F}_{\mu\nu}\mathcal{F}^{\mu\nu}
=-\frac{1}{4g^{2}}\int d^{4}x\,(F^{a}_{\mu\nu}F^{a\mu\nu}+F^{i}_{\mu\nu}F^{i\mu\nu})\,,
\label{YM_action}
\end{equation}
where $g$ is the coupling constant. Furthermore, use has been made of the trace relations
\begin{equation}
\mathrm{tr}(T^{a}T^{b})=\frac{1}{2}\delta^{ab}\,,\qquad
\mathrm{tr}(T^{i}T^{j})=\frac{1}{2}\delta^{ij}\,,\qquad
\mathrm{tr}(T^{a}T^{i})=0\,.
\end{equation}
It is straightforward to show that the action \eqref{YM_action} is left invariant  under the infinitesimal gauge transformation
\begin{eqnarray}
\delta A^{i}_{\mu}&=&-\left(\partial_{\mu}\omega^{i}+f^{abi}A^{a}_{\mu}\omega^{b}\right)\,,\nonumber\\
\delta A^{a}_{\mu}&=&-\left(D^{ab}_{\mu}\omega^{b}+f^{abc}A^{b}_{\mu}\omega^{c}+f^{abi}A^{b}_{\mu}\omega^{i}\right)\,,
\label{gauge_transf}
\end{eqnarray}
where $\omega(x)=\omega^{i}(x)T^{i}+\omega^{a}(x)T^{a}$ is the local gauge parameter, which is taken to be infinitesimal.
\subsection{Defining the ordinary MAG}
Once we have displayed the decomposition of the gauge group in its diagonal and off-diagonal components, we can now open the possibility of choose a gauge-fixing condition in such a way that the diagonal and off-diagonal components can be fixed separately, {\it i.e.} one can define a gauge-fixing condition for
the diagonal sector followed by a different gauge-fixing condition for the off-diagonal sector. This is the case of the MAG, which is defined as
\begin{eqnarray}
D^{ab}_{\mu}A^{b\mu}&=&0\,,\label{MAG1}\\
\partial_{\mu}A^{i\mu}&=&0\,.
\label{MAG2}
\end{eqnarray}
The off-diagonal gauge condition, eq.~\eqref{MAG1}, is obtained by taking the minimum (or, being more rigorous, the extremum condition), under the variations
\eqref{gauge_transf}, of the auxiliary functional
\begin{equation}
\mathcal{H}[A]=\frac{1}{2}\int d^{4}x\, A^{a}_{\mu}A^{a\mu}\,.
\end{equation}
The existence of such auxiliary functional permits the definition of the MAG on the lattice (such as the Landau gauge). The diagonal gauge condition,
eq.~\eqref{MAG2}, cannot be obtained from the minimum condition of any known auxiliary functional. However, a gauge-fixing condition for the diagonal
sector of the theory is necessary due to the residual U$(1)^{n-1}$ symmetry of the Abelian Cartan subgroup. Then, eq.~\eqref{MAG2} is the easiest gauge
condition that can be chosen for the diagonal sector\footnote{There is in the literature a different choice of the diagonal gauge fixing condition. This
gives rise to the so-called modified maximal Abelian gauge (MMAG) \cite{Dudal:2005zr}.}.\\\\Now, following the Faddeev-Popov gauge-fixing procedure, we can write the generating
functional of the Yang-Mills theory gauge-fixed in the MAG as
\begin{eqnarray}
Z_{\mathrm{MAG}}[J,K,\eta,\eta_{\star}]&\propto& \int \mathrm{D}A\,\mathrm{D}b\,\mathrm{D}c\,\mathrm{D}c_{\star}\,
\exp\left[iS_{\mathrm{MAG}}[A,b,c,c_{\star}]+i\int d^{4}x\,(A^{a}_{\mu}J^{a\mu}
+b^{a}K^{a}+\eta_{\star}^{a}c^{a}+c^{a}_{\star}\eta^{a})\right]\,,\nonumber\\
S_{\mathrm{MAG}}[A,b,c,c_{\star}]&=&S_{\mathrm{YM}}+\int d^{4}x\,\bigg[
b^{a}\,D^{ab}_{\mu}A^{b\mu}
+b^{i}\,\partial_{\mu}A^{i\mu}
+c^{a}_{\star}\,D^{ab}_{\mu}D^{bc\mu}c^{c}
+f^{abi}\,c^{a}_{\star}\left(D^{bc}_{\mu}A^{c\mu}\right)c^{i}\nonumber\\
&&
-f^{abi}f^{cdi}\,c^{a}_{\star}c^{d}A^{b}_{\mu}A^{c\mu}
+c^{a}_{\star}\,D^{ab}_{\mu}(f^{bcd}A^{c\mu}c^{d})
+c^{i}_{\star}\,\partial^{\mu}(\partial_{\mu}c^{i}
+f^{abi}A^{a}_{\mu}c^{b})\bigg]\,.
\label{S_MAG}
\end{eqnarray}
There are some points to be properly clarified in \eqref{S_MAG}. First, we call attention to the pair of fields $b^{a}$ and $b^{i}$, known as the
Nakanishi-Lautrup auxiliary fields. Here, these fields have the function of Lagrange multipliers, enforcing the MAG conditions \eqref{MAG1} and \eqref{MAG2},
as one can easily check from their equations of motion,
\begin{equation}
\frac{\delta S_{\mathrm{MAG}}}{\delta b^{a}}= D^{ab}_{\mu}A^{b\mu}\,,\qquad
\frac{\delta S_{\mathrm{MAG}}}{\delta b^{i}}= \partial_{\mu}A^{i\mu}\,.
\end{equation}
Second, the two pairs of complex Grassmannian fields $\{c^{a},c^{a}_{\star}\}$ and $\{c^{i},c^{i}_{\star}\}$ are, respectively, the off-diagonal and diagonal
Faddeev-Popov ghost fields (with the subscript ``$\star$" designating the anti-ghost field). Notice that due to the non-linearity of the MAG there are more
interaction vertices than in the Landau or linear covariant gauges. In fact, there are vertices involving two gluons and two ghosts. These vertices give rise to
additional four ghost interactions that might be taken into account in the renormalization procedure. However, we shall not consider such additional terms in the
present discussion (for a detailed analysis on this subject we refer \cite{Fazio:2001rm}). Third, the classical fields $\{J^{a}_{\mu}, K^{a},\eta^{a},\eta^{a}_{\star}\}$ are
interpreted here as external sources. They are important in order to obtain the Green functions from the generating functional then they can, from now on, be
taken to zero. Finally, it is straightforward to check that the action $S_{\mathrm{MAG}}$ is left invariant by the following set of nilpotent BRST transformations:
\begin{equation}
\begin{tabular}{lll}
$sA^{a}_{\mu}=-(D^{ab}_{\mu}c^{b}+f^{abc}A^{b}_{\mu}c^{c}+f^{abi}A^{b}_{\mu}c^{i})\,,\qquad$
&$sA^{i}_{\mu}=-(\partial_{\mu}c^{i}+f^{abi}A^{a}_{\mu}c^{b})\,,$&$\!\phantom{\Big|}$\cr
$\displaystyle sc^{a}=f^{abi}c^{b}c^{i}+\frac{1}{2}f^{abc}c^{b}c^{c}\,,$
&$\displaystyle sc^{i}=\frac{1}{2}f^{abi}c^{a}c^{b}\,,$&$\!\phantom{\Big|}$\cr
$sc^{a}_{\star}=b^{a}\,,$&$sc^{i}_{\star}=b^{i}
\,,$&$\!\phantom{\Big|}$\cr
$sb^{a}=0\,,$&$sb^{i}=0\,.$&$\!\phantom{\Big|}$\cr
\end{tabular}
\end{equation}
\subsection{The $N=1$ supersymmetric generalization of the MAG}
Our task now is to generalize the MAG, defined by eqs \eqref{MAG1} and \eqref{MAG2}, in the case of the $N=1$ super-Yang-Mills in the formalim of superfields.
In order to achieve this aim, our starting point will be the superfield, $\Phi(x,\theta,\bar\theta)$, which can be written, taking the notations of
\cite{Quevedo:2010ui,Ryder:1985wq,Piguet:1996ys}, as\footnote{In eq.\eqref{superfield} we have: $\theta^{\alpha}$ $(\alpha=1,2)$ and $\bar\theta_{\dot{\alpha}}$
$(\dot{\alpha}=\dot{1},\dot{2})$ being the fermionic coordinates of the superspace;
$\theta^{2}=\theta^{\alpha}\theta_{\alpha}=\varepsilon^{\alpha\beta}\theta_{\beta}\theta_{\alpha}$ and
$\bar\theta^{2}=\bar\theta_{\dot\alpha}\bar\theta^{\dot\alpha}=\varepsilon_{\dot{\alpha}\dot{\beta}}\bar\theta^{\dot{\beta}}\bar\theta^{\dot{\alpha}}$,
with $\varepsilon^{12}=-\varepsilon^{21}=-\varepsilon_{\dot{1}\dot{2}}=\varepsilon_{\dot{2}\dot{1}}=1$ and
$\varepsilon^{11}=\varepsilon^{22}=\varepsilon_{\dot{1}\dot{1}}=\varepsilon_{\dot{2}\dot{2}}=0$; and also
$$
[(\sigma^{0})_{\alpha\dot\alpha}]=
\left(\begin{matrix}
1&0\cr
0&1
\end{matrix}\right)\,,\qquad
[(\sigma^{1})_{\alpha\dot\alpha}]=
\left(\begin{matrix}
0&1\cr
1&0
\end{matrix}\right)\,,\qquad
[(\sigma^{2})_{\alpha\dot\alpha}]=
\left(\begin{matrix}
0&-i\cr
i&0
\end{matrix}\right)\,,\qquad
[(\sigma^{3})_{\alpha\dot\alpha}]=
\left(\begin{matrix}
1&0\cr
0&-1
\end{matrix}\right)\,.
$$
}
\begin{eqnarray}
\Phi(x,\theta,\bar\theta)&=&\Phi^{A}(x,\theta,\bar\theta)T^{A}\nonumber\\
\Phi^{A}(x,\theta,\bar\theta)&=&C^{A}(x)+\theta^{\alpha}\chi^{A}_{\alpha}(x)-\bar\theta^{\dot{\alpha}}\bar\chi^{A}_{\dot{\alpha}}(x)
+\frac{1}{2}\,\theta^{2}M^{A}(x)+\frac{1}{2}\,\bar\theta^{2}\bar{M}^{A}(x)\nonumber\\
&&+2\,\theta^{\alpha}(\sigma^{\mu})_{\alpha\dot{\alpha}}\bar\theta^{\dot{\alpha}}\,A^{A}_{\mu}(x)
+\frac{1}{2}\,\bar\theta^{2}\theta^{\alpha}\lambda^{A}_{\alpha}(x)
-\frac{1}{2}\,\theta^{2}\bar\theta^{\dot{\alpha}}\bar\lambda^{A}_{\dot{\alpha}}(x)
+\frac{1}{4}\,\theta^{2}\bar\theta^{2}\,\mathfrak{D}^{A}(x)\,.
\label{superfield}
\end{eqnarray}
Here, $\{C,\chi_{{\alpha}},\bar\chi_{\dot{\alpha}},M,\bar{M},A_{\mu},\lambda_{\alpha},\bar\lambda_{\dot\alpha},\mathfrak{D}\}$ are the components of the superfield
taking values in the adjoint representation of the SU($n$) group. Thus, the strategy that we will adopt can be summarized in three steps as
\begin{itemize}
\item{Impose gauge conditions on the diagonal and
off-diagonal components of the superfield $\{\Phi^{i},\Phi^{a}\}$;}
\item{Impose the Wess-Zumino gauge, {\it i.e.} take $\{C,\chi_{{\alpha}},\bar\chi_{\dot{\alpha}},M,\bar{M}\}=0$;}
\item{Verify if the vectorial degrees of freedom of the supermultiplet obey the MAG conditions
\eqref{MAG1} and \eqref{MAG2}.}
\end{itemize}
Notice that the second step indicates that the original MAG can be recovered only together with the Wess-Zumino gauge. However, this step is not necessary, for
example, in the Landau gauge, as one can verify in \cite{Piguet:1996ys}. This necessity could be an effect of the nonlinearity of the MAG.\\\\Before applying the
three steps listed above, let us take a look at the following quantity:
\begin{equation}
\bar{\mathcal{D}}^{2}\mathcal{D}^{2}\Phi^{A} =e^{i(\theta\sigma^{\mu}\bar\theta)\partial_{\mu}}\,\left[
4\left(\mathfrak{D}^{A}-\partial^{2}C^{A}+4i\,\partial_{\nu}A^{A\nu}\right)
+8i\theta^{\alpha}(\sigma^{\nu})_{\alpha\dot\alpha}\,\partial_{\nu}\left(\bar\lambda^{A\dot\alpha}
-i(\sigma^{\rho})^{\phantom{\alpha}\dot\alpha}_{\beta}\,\partial_{\rho}\chi^{A\beta}\right)
-8\theta^{2}\partial^{2}M^{A}\right]\,,
\label{superLandau}
\end{equation}
where the covariant derivatives $\mathcal{D}_{\alpha}$ and $\bar{\mathcal{D}}_{\dot\alpha}$ are, following \cite{Quevedo:2010ui}, given by
\begin{equation}
\mathcal{D}_{\alpha}=\frac{\partial}{\partial\theta^{\alpha}}+i(\sigma^{\mu})_{\alpha\dot\alpha}\,\bar\theta^{\dot\alpha}\partial_{\mu}\,,\qquad
\bar{\mathcal{D}}_{\dot\alpha}=-\frac{\partial}{\partial\bar\theta^{\dot\alpha}}-i\theta^{\alpha}(\sigma^{\mu})_{\alpha\dot\alpha}\,\partial_{\mu}\,.
\end{equation}
This quantity is clearly a chiral superfield, {\it i.e.} $\bar{\mathcal{D}}_{\dot\alpha}(\bar{\mathcal{D}}^{2}\mathcal{D}^{2}\Phi^{A})=0$, and it contains, as
one of its components, the 4-divergence of the vectorial component of $\Phi(x,\theta,\bar\theta)$. Then, the supersymmetric extension of the Landau gauge, for
example, can be achieved by demanding that this quantity vanishes, {\it i.e.} $\bar{\mathcal{D}}^{2}\mathcal{D}^{2}\Phi^{A}=0$, see \cite{Piguet:1996ys}
\footnote{The notation in \cite{Piguet:1996ys} is slightly different of the one presented here. Then, in order to verify that the result obtained in
\cite{Piguet:1996ys}
is in accordance
with that of \eqref{superLandau}, it is necessary to perform the replacement $(i\to -i)$ in eq.~\eqref{superLandau}.}.
As stated in the previous section, the gauge-fixing condition of the diagonal component of the gauge field, eq.~\eqref{MAG2}, is entirely analogous to the Landau
gauge. Thus, for the diagonal component of the superfield $\Phi(x,\theta,\bar\theta)$, we can choose, as a supersymmetric generalization of \eqref{MAG2}, the
following gauge condition:
\begin{equation}
\bar{\mathcal{D}}^{2}\mathcal{D}^{2}\Phi^{i}=0\,.
\label{diag_super_MAG}
\end{equation}
Now, we have to define a gauge-fixing condition on the off-diagonal components of the superfield in such a way to recover the nonlinear MAG condition \eqref{MAG1}.
In fact, eq.~\eqref{MAG1} can be explicitly written as
\begin{equation}
\partial_{\mu}A^{a\mu}=f^{abi}A^{i}_{\mu}A^{b\mu}\,.
\label{MAG_again}
\end{equation}
One can immediately observe that the r.h.s. of the equation above is nonlinear. Thus, our task is to find a gauge condition for $\Phi^{a}(x,\theta,\bar\theta)$,
the off-diagonal component of the superfield, which contains $f^{abi}A^{i}_{\mu}A^{b\mu}$. The l.h.s. of \eqref{MAG_again} is already contained in
$\bar{\mathcal{D}}^{2}\mathcal{D}^{2}\Phi^{a}$, as one can see in \eqref{superLandau}. Then, the gauge condition we are looking for might be of the type
\begin{equation}
\bar{\mathcal{D}}^{2}\mathcal{D}^{2}\Phi^{a}+\Psi^{a}(x,\theta,\bar\theta)=0\,,
\label{pre_super_MAG}
\end{equation}
where $f^{abi}A^{i}_{\mu}A^{b\mu}\in \Psi^{a}(x,\theta,\bar\theta)$. To be more precise, we would like that
\begin{equation}
\Psi^{a}(x,\theta,\bar\theta)=e^{i(\theta\sigma^{\mu}\bar\theta)\partial_{\mu}}\,\left(-16if^{abi}A^{i}_{\nu}A^{b\nu}+u^{a}(x)
+\theta^{\alpha}v^{a}_{\alpha}(x)+\theta^{2}w^{a}(x)\right)\,,
\label{psi}
\end{equation}
where $u^{a}(x)$, $v^{a}_{\alpha}(x)$ and $w^{a}(x)$ are, at this point, general functions that will be determined later when we  precisely define an expression for $\Psi^{a}(x,\theta,\bar\theta)$. Therefore, taking into account \eqref{psi}, a possible candidate to represent the desired
$\Psi^{a}(x,\theta,\bar\theta)$ is
\begin{equation}
\Psi^{a}(x,\theta,\bar\theta)=-\frac{i}{2}\bar{\mathcal{D}}^{2}\mathcal{D}^{2}\left(f^{abi}\Phi^{i}\Phi^{b}\right)\,.
\label{comparing}
\end{equation}
Then,
\begin{eqnarray}
u^{a}(x)&=&-2if^{abi}\Big[C^{i}\mathfrak{D}^{b}+\mathfrak{D}^{i}C^{b}-\chi^{i\alpha}\lambda^{b}_{\alpha}
-\lambda^{i\alpha}\chi^{b}_{\alpha}
+\bar\chi^{i\dot\alpha}\bar\lambda^{b}_{\dot\alpha}
+\bar\lambda^{i\dot\alpha}\bar\chi^{b}_{\dot\alpha}
+M^{i}\bar{M}^{b}
+\bar{M}^{i}M^{b}
-\partial^{2}\big(C^{i}C^{b}\big)\nonumber\\
&&
+i\partial_{\mu}\left(4C^{i}A^{b\mu}+4A^{i\mu}C^{b}
+\chi^{i}\sigma^{\mu}\bar\chi^{b}
+\chi^{b}\sigma^{\mu}\bar\chi^{i}\right)\Big]\,,\\
v^{a}_{\alpha}(x)&=&4if^{abi}\Big[\partial^{2}\left(C^{i}\chi^{b}_{\alpha}+\chi^{i}_{\alpha}C^{b}\right)
-i(\sigma^{\mu})_{\alpha\dot\alpha}\,\partial_{\mu}\Big(C^{i}\bar\lambda^{b\dot\alpha}
+\bar\lambda^{i\dot\alpha}C^{b}
+2\bar\chi^{i\dot\alpha}M^{b}
+2M^{i}\bar\chi^{b\alpha}\nonumber\\
&&+2(\sigma^{\nu})_{\beta}^{\phantom{\alpha}\dot\alpha}\,\chi^{i\beta}A^{b}_{\nu}
+2(\sigma^{\nu})_{\beta}^{\phantom{\alpha}\dot\alpha}\,A^{i}_{\nu}\chi^{b\beta}\Big)\Big]\,,\\
w^{a}(x)&=&2if^{abi}\partial^{2}\left(
C^{i}M^{b}+M^{i}C^{b}+\chi^{i}_{\alpha}\chi^{b\alpha}\right)\,.
\end{eqnarray}
Collecting the results obtained from eqs.~\eqref{diag_super_MAG}, \eqref{pre_super_MAG} and \eqref{comparing}, we are in a position to define the supersymmetric
extension of the MAG as
\begin{eqnarray}
\bar{\mathcal{D}}^{2}\mathcal{D}^{2}\left(\Phi^{a}-\frac{i}{2}
f^{abi}\Phi^{i}\Phi^{b}\right)&=&0\,,
\label{superMAG1}\\
\bar{\mathcal{D}}^{2}\mathcal{D}^{2}\Phi^{i}&=&0\,.
\label{superMAG2}
\end{eqnarray}
Notice that the gauge conditions \eqref{superMAG1} and \eqref{superMAG2} are taken independently of the Wess-Zumino gauge. Once we have established the gauge conditions on the components of the superfield, the second step is to take the Wess-Zumino gauge in order to confirm that the original MAG can be re-obtained. In our case, it is sufficient to verify the off-diagonal gauge condition \eqref{superMAG1}:
\begin{equation}
\left.\bar{\mathcal{D}}^{2}\mathcal{D}^{2}\left(\Phi^{a}-\frac{i}{2}
f^{abi}\Phi^{i}\Phi^{b}\right)\right|_{\mathrm{WZ}}=e^{i(\theta\sigma^{\mu}\bar\theta)\partial_{\mu}}\,\left[
4\mathfrak{D}^{a}+16i\,(\partial_{\nu}A^{a\nu}-f^{abi}A^{i}_{\nu}A^{b\nu})
+8i\theta^{\alpha}(\sigma^{\nu})_{\alpha\dot\alpha}\,\partial_{\nu}\bar\lambda^{a\dot\alpha}
\right]\,.
\end{equation}
In other words, we can safely say that the MAG is a limit case of the gauge defined by the conditions \eqref{superMAG1} and \eqref{superMAG2} when the Wess-Zumino gauge is taken.

\section{The Faddeev-Popov quantization, BRST symmetry and the Slavnov-Taylor Identity}
Let us start this section by briefly presenting the well-known (pure) SYM action and its gauge invariance. Taking the superfield $\Phi(x,\theta,\bar\theta)$, previously defined in  eq.~\eqref{superfield}, one can construct the supersymmetric generalization of the field strength $\mathcal{F}_{\mu\nu}$ as
\begin{equation}
\mathcal{W}_{\alpha}=W^{A}_{\alpha}T^{A}=\bar{\mathcal{D}}^{2}(e^{-\Phi}\mathcal{D}_{\alpha}e^{\Phi})\,.
\end{equation}
Then, the SYM action will be given by
\begin{equation}
S_{\mathrm{SYM}}=-\frac{1}{64g^{2}}\mathrm{tr}\int d^{4}xd^{2}\theta\,\mathcal{W}^{\alpha}\mathcal{W}_{\alpha}
=-\frac{1}{128g^{2}}\int d^{4}xd^{2}\theta\,W^{A\alpha}W^{A}_{\alpha}\,.
\label{SYM_action}
\end{equation}
This action is left invariant under the following infinitesimal gauge transformation (for details see \cite{Piguet:1996ys}):
\begin{eqnarray}
\Phi&\to&\Phi'=\Phi+\delta\Phi\,,\nonumber\\
\delta\Phi&=&\frac{i}{2}L_\Phi(\Lambda+\bar\Lambda)+\frac{i}{2}\left[L_\Phi\,\mathrm{coth}\left(\frac{1}{2}L_\Phi\right)\right]
(\Lambda-\bar\Lambda)\nonumber\\
&=&i(\Lambda-\bar\Lambda)
+\frac{1}{2}[\Phi,\Lambda+\bar\Lambda]+\frac{i}{12}[\Phi,[\Phi,\Lambda-\bar\Lambda]]+\mathcal{O}(\Phi^{3})\,,
\end{eqnarray}
where $L_{\Phi}\bullet=[\Phi,\bullet]$ and $\Lambda=\Lambda^{A}T^{A}$ is an infinitesimal chiral superfield, while $\bar\Lambda=\bar\Lambda^{A}T^{A}$ is an anti-chiral one. As one can notice, the gauge invariance of \eqref{SYM_action} lead us to perform a gauge-fixing procedure in order correct quantize the theory. Thus, we follow \cite{Piguet:1996ys} in order to establish the Faddeev-Popov quantization approach assuming the supersymmetric MAG, defined by eq's~\eqref{superMAG1} and \eqref{superMAG2}, as the gauge-fixing. We first notice that the conditions \eqref{superMAG1} and
\eqref{superMAG2} might be introduced in the theory with the help of Lagrange multipliers, in the same way we have introduced the Nakanishi-Lautrup auxiliary fields $\{b^{a},b^{i}\}$ in the case of the ordinary MAG. Then, let us consider the gauge-fixing term
\begin{eqnarray}
S_{\mathrm{gf}}&=&
\frac{1}{8}\int d^{4}xd^{2}\theta\left[B^{a}\bar{\mathcal{D}}^{2}\mathcal{D}^{2}\left(\Phi^{a}-\frac{i}{2}
f^{abi}\Phi^{i}\Phi^{b}\right)+B^{i}\bar{\mathcal{D}}^{2}\mathcal{D}^{2}\Phi^{i}\right]+\mathrm{c.c.}\nonumber\\
&=&\frac{1}{8}\int dV\left[B^{a}\mathcal{D}^{2}\left(\Phi^{a}-\frac{i}{2}
f^{abi}\Phi^{i}\Phi^{b}\right)+B^{i}\mathcal{D}^{2}\Phi^{i}
+\bar{B}^{a}\bar{\mathcal{D}}^{2}\left(\Phi^{a}+\frac{i}{2}
f^{abi}\Phi^{i}\Phi^{b}\right)+\bar{B}^{i}\bar{\mathcal{D}}^{2}\Phi^{i}\right]\,,
\label{gf}
\end{eqnarray}
where $dV=d^{4}xd^{2}\theta d^{2}\bar\theta$ is the superspace element volume, and, playing the role of Lagrange multipliers, the pairs $\{B^{a},B^{i}\}$ and $\{\bar{B}^{a},\bar{B}^{i}\}$ of chiral and anti-chiral superfields, respectively. Now, we need to find a BRST invariant way to add the gauge-fixing term \eqref{gf} term to the SYM action. In other words, we have to introduce the Faddeev-Popov ghost fields. As the gauge transformation involves the chiral and anti-chiral parameters $\{\Lambda,\bar\Lambda\}$, and taking into account that the gauge-fixing separates the diagonal and off-diagonal components of the symmetry group we have to deal with several ``types" of ghost fields:
\begin{itemize}
\item{The off-diagonal chiral ghosts: $\{c^{a},c^{a}_{\star}\};$}
\item{The off-diagonal anti-chiral ghosts: $\{\bar{c}^{a},\bar{c}^{a}_{\star}\};$}
\item{The diagonal chiral ghosts: $\{c^{i},c^{i}_{\star}\};$}
\item{The diagonal anti-chiral ghosts: $\{\bar{c}^{a},\bar{c}^{a}_{\star}\}.$}
\end{itemize}
With this field content at our disposal, we display below a full set of nilpotent $(s^{2}=0)$ BRST transformations under which we demand that the complete gauge-fixed action be invariant:
\begin{itemize}
\item{Transformations of the components of the superfield $\Phi(x,\theta,\bar\theta)$:
\begin{eqnarray}
s\Phi^{a}&=&i(c^{a}-\bar{c}^{a})
-\frac{1}{2}f^{abc}\Phi^{b}(c^{c}+\bar{c}^{c})
-\frac{1}{2}f^{abi}\Phi^{b}(c^{i}+\bar{c}^{i})
+\frac{1}{2}f^{abc}\Phi^{i}(c^{b}+\bar{c}^{b})
+\mathcal{O}(\Phi^{2})\,,\nonumber\\
s\Phi^{i}&=&i(c^{i}-\bar{c}^{i})
-\frac{1}{2}f^{abi}\Phi^{a}(c^{b}+\bar{c}^{b})
+\mathcal{O}(\Phi^{2})\,;
\label{brst1}
\end{eqnarray}}
\item{Transformations of the components of chiral superfields $\{c,c_{\star},B\}$:
\begin{equation}
\begin{tabular}{lll}
$\displaystyle sc^{a}=f^{abi}c^{b}c^{i}+\frac{1}{2}f^{abc}c^{b}c^{c}\,,\qquad$
&$\displaystyle sc^{i}=\frac{1}{2}f^{abi}c^{a}c^{b}\,,$&$\!\phantom{\Big|}$\cr
$sc^{a}_{\star}=B^{a}\,,$&$sc^{i}_{\star}=B^{i}
\,,$&$\!\phantom{\Big|}$\cr
$sB^{a}=0\,,$&$sB^{i}=0\,;$&$\!\phantom{\Big|}$\cr
\end{tabular}
\label{brst2}
\end{equation}}
\item{Transformations of the components of the anti-chiral superfields $\{\bar{c},\bar{c}_{\star},\bar{B}\}$:
\begin{equation}
\begin{tabular}{lll}
$\displaystyle s\bar{c}^{a}=f^{abi}\bar{c}^{b}\bar{c}^{i}+\frac{1}{2}f^{abc}\bar{c}^{b}\bar{c}^{c}\,,\qquad$
&$\displaystyle s\bar{c}^{i}=\frac{1}{2}f^{abi}\bar{c}^{a}\bar{c}^{b}\,,$&$\!\phantom{\Big|}$\cr
$s\bar{c}^{a}_{\star}=\bar{B}^{a}\,,$&$s\bar{c}^{i}_{\star}=\bar{B}^{i}
\,,$&$\!\phantom{\Big|}$\cr
$s\bar{B}^{a}=0\,,$&$s\bar{B}^{i}=0\,.$&$\!\phantom{\Big|}$\cr
\end{tabular}
\label{brst3}
\end{equation}}
\end{itemize}
Thus, we can define the BRST invariant SYM action gauge-fixed in the supersymmetric MAG:
\begin{eqnarray}
S&:=&S_{\mathrm{SYM}}+\frac{1}{8}s\int dV\left[c^{a}_{\star}\mathcal{D}^{2}\left(\Phi^{a}-\frac{i}{2}f^{abi}\Phi^{i}\Phi^{b}\right)
+c^{i}_{\star}\mathcal{D}^{2}\Phi^{i}
+\bar{c}^{a}_{\star}\bar{\mathcal{D}}^{2}\left(\Phi^{a}+\frac{i}{2}f^{abi}\Phi^{i}\Phi^{b}\right)
+\bar{c}^{i}_{\star}\bar{\mathcal{D}}^{2}\Phi^{i}
\right]\nonumber\\
&=&S_{\mathrm{SYM}}+\frac{1}{8}\int dV\left[B^{a}\mathcal{D}^{2}\left(\Phi^{a}-\frac{i}{2}f^{abi}\Phi^{i}\Phi^{b}\right)
+B^{i}\mathcal{D}^{2}\Phi^{i}
+\bar{B}^{a}\bar{\mathcal{D}}^{2}\left(\Phi^{a}+\frac{i}{2}f^{abi}\Phi^{i}\Phi^{b}\right)
+\bar{B}^{i}\bar{\mathcal{D}}^{2}\Phi^{i}\right]
\nonumber\\
&&-\frac{1}{8}\int dV\left[c^{a}_{\star}\mathcal{D}^{2}s\left(\Phi^{a}-\frac{i}{2}f^{abi}\Phi^{i}\Phi^{b}\right)
+c^{i}_{\star}\mathcal{D}^{2}s\Phi^{i}
+\bar{c}^{a}_{\star}\bar{\mathcal{D}}^{2}s\left(\Phi^{a}+\frac{i}{2}f^{abi}\Phi^{i}\Phi^{b}\right)
+\bar{c}^{i}_{\star}\bar{\mathcal{D}}^{2}s\Phi^{i}\right]\,.
\label{SYM+MAG}
\end{eqnarray}
The BRST symmetry of the theory can be expressed as a Ward identity. However, as one can notice in eq's~\eqref{brst1}, \eqref{brst2} and \eqref{brst3}, there are several nonlinear BRST transformations. These nonlinear transformations must be introduced in the theory as composite operators \cite{Piguet:1995er,Piguet:1996ys}. It is important in order to establish the renormalization of the symmetry itself. Therefore, it is necessary to take into account these nonlinearities by coupling them with external sources and adding to the action $S$ the term
\begin{equation}
S_{\mathrm{ext}}=\int dV \left(-\Omega^{a}s\Phi^{a}-\Omega^{i}s\Phi^{i}\right)
+\int d^{4}x d^{2}\theta\left(L^{a}sc^{a}+L^{i}sc^{i}\right)
+\int d^{4}x d^{2}\bar\theta\left(\bar{L}^{a}s\bar{c}^{a}+\bar{L}^{i}s\bar{c}^{i}\right)\,.
\label{ext}
\end{equation}
The sources $\{\Omega,L,\bar{L}\}$ are taken as classical superfields, with $\Omega^{a,i}$ being anti-commuting, or Grassmannian, variables and $\{L^{a,i},\bar{L}^{a,i}\}$ being commuting ones. The BRST invariance of $S_{\mathrm{ext}}$ is obtained by demanding that
\begin{equation}
s\{\Omega,L,\bar{L}\}=0\,.
\end{equation}
Then, we can finally define our complete BRST invariant action as
\begin{equation}
\Sigma:=S+S_{\mathrm{ext}}\,,
\label{full_action}
\end{equation}
with $S$ and $S_{\mathrm{ext}}$ given, respectively, by \eqref{SYM+MAG} and \eqref{ext}. Now, we are able to write the Slavnov-Taylor identity, which represents the BRST symmetry as a Ward identity. Namely,
\begin{eqnarray}
\mathcal{S}(\Sigma)&:=&\int dV\left(
\frac{\delta\Sigma}{\delta\Omega^{a}}\frac{\delta\Sigma}{\delta\Phi^{a}}
+\frac{\delta\Sigma}{\delta\Omega^{i}}\frac{\delta\Sigma}{\delta\Phi^{i}}\right)
+\int d^{4}x d^{2}\theta\left(\frac{\delta\Sigma}{\delta L^{a}}\frac{\delta\Sigma}{\delta c^{a}}
+\frac{\delta\Sigma}{\delta L^{i}}\frac{\delta\Sigma}{\delta c^{i}}
+B^{a}\frac{\delta\Sigma}{\delta c^{a}_{\star}}
+B^{i}\frac{\delta\Sigma}{\delta c^{i}_{\star}}\right)\nonumber\\
&&
+\int d^{4}xd^{2}\bar\theta\left(\frac{\delta\Sigma}{\delta \bar{L}^{a}}\frac{\delta\Sigma}{\delta \bar{c}^{a}}
+\frac{\delta\Sigma}{\delta \bar{L}^{i}}\frac{\delta\Sigma}{\delta \bar{c}^{i}}
+\bar{B}^{a}\frac{\delta\Sigma}{\delta \bar{c}^{a}_{\star}}
+\bar{B}^{i}\frac{\delta\Sigma}{\delta \bar{c}^{i}_{\star}}\right)=0\,.
\label{ST}
\end{eqnarray}
It is worth mentioning that there are also important Ward identities that are useful to prove renormalization such as the diagonal gauge-fixing equations and the diagonal anti-ghost equations:
\begin{equation}
\frac{\delta\Sigma}{\delta B^{i}}=\bar{\mathcal{D}}^{2}\mathcal{D}^{2}\Phi^{i}\,,\qquad
\frac{\delta\Sigma}{\delta \bar{B}^{i}}=\mathcal{D}^{2}\bar{\mathcal{D}}^{2}\Phi^{i}\,,
\end{equation}
\begin{equation}
\frac{\delta\Sigma}{\delta c^{i}_{\star}}+\frac{1}{8}\bar{\mathcal{D}}^{2}\mathcal{D}^{2}\frac{\delta\Sigma}{\delta\Omega^{i}}=0\,,\qquad
\frac{\delta\Sigma}{\delta \bar{c}^{i}_{\star}}+\frac{1}{8}{\mathcal{D}}^{2}\bar{\mathcal{D}}^{2}\frac{\delta\Sigma}{\delta\Omega^{i}}=0\,.
\end{equation}

\section{Conclusions and further perspectives}
In this work we have presented a supersymmetric generalization of the MAG. In other words, the ordinary gauge conditions \eqref{MAG1} and \eqref{MAG2}, which define the MAG, are generalized to the gauge conditions \eqref{superMAG1} and \eqref{superMAG2} imposed directly on the off-diagonal and diagonal components of the superfield, respectively, and the original MAG is then reobtained from its supersymmetric version when the Wess-Zumino gauge is taken. We also presented the Faddeev-Popov quantization approach, obtaining a BRST invariant action \eqref{full_action}. This property of invariance over the BRST transformations allowed us to write the Slavnov-Taylor Ward identity, eq.~\eqref{ST}, which is fundamental in the study of the renormalizability of the action \eqref{full_action}. These results open the possibility of future works, which we list below:
\begin{itemize}
\item{The proof of the renormalization of the action \eqref{full_action} will certainly be one of our main concerns. However, it is important to say that it is not a trivial task due to the fact that there are also extra quartic ghost interaction terms that need to be considered. These terms are well-known in the ordinary MAG \cite{Fazio:2001rm}, being a direct consequence of the non-linearity of this gauge.}
\item{Once we have established the supersymmetric MAG, we can immediately study the supersymmetric generalization of other gauges as the modified MAG (MMAG) and the interpolating gauge between the Landau gauge and the MAG \cite{Dudal:2005zr}.}
\item{The Euclidean formulation and the study of the Gribov ambiguity are also one of the subjects to be investigated in the future and a possible generalization of the Gribov-Zwanziger (GZ) model, and its refined version, the RGZ model, could be obtained, as already done in the Landau gauge \cite{Amaral:2013uya,Capri:2014xea}. It is worth mentioning, that the GZ and the RGZ models were extensively studied in the ordinary MAG \cite{Capri:2010an,Capri:2008ak}. Also, as pointed out in \cite{Bruckmann:2000ay,bruckmann}, there are instanton configurations lying on the Gribov horizon in the case of SU(2) ordinary MAG. It might be interesting to investigate if this survives with fermions, in view of the fermionic zero mode of the instanton, due to the index theorem \cite{Vandoren:2008xg}. }
\item{Another investigation which we intend to pursue is the extension of
the present study of the MAG to go beyond simple supersymmetry.
The consideration of the $N=2$ and $N=4$ extended super-Yang-Mills models
would be interesting in view of the ultraviolet finiteness of the $N=4$ case
in general and the specific $N=2$ models which can also be built up to be
finite. For example, a specially interesting issue in the case of $N=4$, the
global SU(4) automorphism of the supersymmetry algebra, which is an
important ingredient to understand its ultraviolet finiteness, should be
explicitly broken if MAG is chosen with the $N=4$ model described in terms
of $N=1$ superfields. So, reassessing $N=4$ finiteness in the framework of the
supersymmetric MAG is an issue of interest and we shall be working to
report on that in a future paper.}
\end{itemize}

\section*{Acknowledgments}
The Conselho Nacional de Desenvolvimento Cient\'{\i}fico e
Tecnol\'{o}gico (CNPq-Brazil), the Faperj, Funda{\c{c}}{\~{a}}o de
Amparo {\`{a}} Pesquisa do Estado do Rio de Janeiro,  the
Coordena{\c{c}}{\~{a}}o de Aperfei{\c{c}}oamento de Pessoal de
N{\'{\i}}vel Superior (CAPES)  are gratefully acknowledged. We also would like to thank Markus Q. Huber for nice discussions at CBPF in an early stage of this work.

\end{document}